\documentstyle[aps,prl,floats,graphicx]{revtex}

\begin{document}

\def\nuc#1#2{${}^{#1}$#2}
\def\per{Pb(ClO$_4$)$_2$}
\def\EOne{$\bar{E_1}$}
\def\ETwo{$\bar{E_2}$}
\def\today{\space\number\day\space\ifcase\month\or January\or February\or
  March\or April\or May\or June\or July\or August\or September\or October\or
  November\or December\fi\space\number\year}

\draft

\twocolumn[\hsize\textwidth\columnwidth\hsize\csname@twocolumnfalse\endcsname

\title{Measuring Supernova Neutrino Temperatures using Lead Perchlorate }

\author{S.\,R.\,Elliott}
\vspace{-5mm}

\address{ Department of Physics, Box 351560, University of Washington, Seattle, Washington 98195, USA}

\date{\today}

\maketitle

\begin{abstract}
        Neutrino interactions
with lead produce neutrons in numbers that depend on neutrino energy and type. A detector based on
lead perchlorate, for example, would be able to measure the energy deposited by electrons
and gammas in coincidence with the number of neutrons produced. Sorting the
electron energy spectra by the number of coincident neutrons permits the identification
of the neutrino type that induced the reaction. This separation
allows an analysis which can determine the 
temperatures of $\nu_e$ and $\bar{\nu}_{e}$ from a supernova in one experiment. 
The neutrino reaction signatures of lead perchlorate and 
 the fundamentals of using this material as a neutrino detector are described.
\end{abstract}

\pacs{PACS numbers: 97.60.Bw, 25.30.Pt}

] 

\section{Introduction}

Recently a number of groups have expressed interest in using Pb as a target for neutrino
interactions to study supernovae\cite{LAND,OMNIS} or oscillations\cite{HAR98}, and this inspired
an estimate of the cross section\cite{LAND}.
 The interest arises because of the large cross section and the 
low relative cost of Pb. As a result, additional cross section calculations were
done recently by Fuller, Haxton, and McLaughlin\cite{FUL99} (hereafter referred to as FHM) and
Kolbe and Langanke\cite{KOL99a,KOL00} (hereafter referred to as KL).

The interesting neutrino interactions with Pb consist of

$
\begin{array}{lclr}
\nu_e + ^{208}Pb & \Rightarrow & ^{208}Bi^* + e^- 	& (CC)\\
                        &             & \Downarrow & \\
                        &             & ^{208-y}Bi + x\gamma + yn & \\
\\
\nu_x + ^{208}Pb & \Rightarrow & ^{208}Pb^* + \nu_{x}^{'} & (NC) \\
                        &             & \Downarrow & \\
                        &             &^{208-y}Pb + x\gamma + yn. & \\
\end{array}
$

\begin{figure}
\begin{center}
\includegraphics[width=2. in]{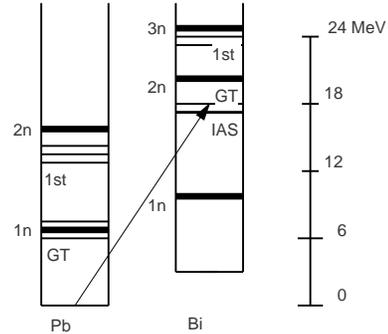}
\end{center}
\vspace{0.5in}
\caption{The level scheme of \nuc{208}{Pb} - \nuc{208}{Bi} system. The levels labeled
with {\it GT} indicating the Gamow-Teller resonance, {\it IAS} indicates the isobaric analog state, and
{\it 1st} indicating the states populated by first forbidden transitions. The  1, 2, and 3 neutron 
emission thresholds are indicated by the labels {\it 1n}, {\it 2n}, and {\it 3n} respectively}
\label{LevelScheme}
\end{figure}

\noindent Fig. \ref{LevelScheme} shows the energetics of these transitions. The number of neutrons 
emitted (0, 1, or 2) depends on the neutrino energy and whether the transition is induced by a
 charged current (CC) or neutral current (NC) interaction. The nuclear physics of this
system is described in detail in Refs. \cite{FUL99,KOL00}.

An ideal lead-based neutrino detector 
would have an appreciable density of lead atoms and the capability
of detecting the electrons, gammas and neutrons produced in the reaction. Lead perchlorate (\per)
has a very high solubility in water (500 g \per / 100 g H$_{2}$O \cite{CRC}) and
the mixture is transparent. This transparency raises the hope that a $\check{C}$erenkov 
detector can be assembled. Additionally, the presence of a neutron moderator (hydrogen) and 
a neutron capture nucleus (\nuc{35}{Cl}) provides a method for observing the neutrons.
 The high-energy (8.6 MeV), neutron-capture $\gamma$ rays from Cl 
would Compton scatter in the fluid and be observed via the $\check{C}$erenkov
light of the recoil electrons.

Supernovas emit on the order 10$^{58}$ $\nu$ of all types.  The average energies of neutrinos emitted by a supernova follow a hierarchy:
 $\overline{E}_{\nu_e} < \overline{E}_{\bar{\nu}_e} < \overline{E}_{\nu_{\mu,\tau}}$ where ${\nu_{\mu,\tau}}$ 
indicates either $\mu$ or
$\tau$ neutrinos and their anti-particles. The predicted values of the average energies for the three neutrino classes differ a great
deal and are estimated to be $\overline{E}_{\nu_e} = 11$ MeV, $\overline{E}_{\bar{\nu}_e} = 16$ MeV, 
and $\overline{E}_{\nu_{\mu,\tau}} = 25$ MeV\cite{MIL93,JAN89}. 

Although not strictly thermal, the supernova neutrino energy spectra can be described
well by a thermal distribution, 

\begin{eqnarray*}
f_{\nu}  & = & \frac{1}{T^3_{\nu}F_2(\eta)}\frac{E^2_{\nu}}{exp(E_{\nu}/T_{\nu} - \eta) + 1} \\
\end{eqnarray*}

\noindent where $f_{\nu}$ is the normalized neutrino spectral shape, $T_{\nu}$ is an effective neutrino temperature,
$E_{\nu}$ is the neutrino energy, and $\eta$ is the degeneracy parameter (chemical potential divided 
by $T_{\nu}$).
For $\nu_e$ and $\bar{\nu}_e$, the supernova model predictions for the flux is well fit with $\eta$ = 3, 
whereas for $\nu_{\mu,\tau}$ the flux is better described with $\eta$ = 0.
The normalization factor $F_2(\eta)$ has the values $F_2(0)$ = 1.803 and $F_2(3)$ = 18.969.  The total neutrino fluence
$F_{\nu}$ from a supernova can be estimated by

\begin{eqnarray*}
F_{\nu}  & = & 2.8 \times 10^{11} cm^{-2} \frac{E_B}{10^{53} erg}\frac{MeV}{T_{\nu}}(\frac{10 kpc}{D})^2 \\
\end{eqnarray*}

\noindent where $E_B$ is the released energy and D is the distance to the supernova.

The supernova $\nu_e$ spectrum is too soft to produce a large number of 
multiple-neutron inverse-beta-decay events in Pb. However, if the higher
energy $\nu_{\mu,\tau}$ oscillate into $\nu_e$, the hardening of the spectrum will greatly alter
the expected number of 2-neutron events\cite{FUL99}. Thus the response of a Pb detector is intriguing because
 if T$_{\nu_e}$ is found to be larger than
T$_{\bar{\nu}_{e}}$, it would be strong evidence that neutrino oscillations are taking place. However, the uncertainties
in the neutrino energy distribution as predicted by the supernova models may complicate the interpretation of a measurement of
the spectral parameters.

The FHM paper demonstrated that the NC and CC interaction
rates in Pb are sensitive probes of $T_{\nu_{\mu},\nu_{\tau}}$, $T_{\nu_e}$, and $T_{\bar{\nu}_e}$. 
This paper further discusses the details of studying supernova neutrinos with Pb.
In particular the use of the added data arising from observing the product electrons and gammas
in coincidence with neutrons is described. These data permit the separation of the NC and CC
rates and a further division of CC $\nu_e$ and ${\bar{\nu}_e}$ events. This collection of data
can be interpreted in terms of the 3 temperatures of interest. The CC event
division is discussed here in detail.
 Although any Pb-based experiment that measures electrons, gammas and neutrons might apply the
techniques described herein, the example of using lead perchlorate as a target material is discussed. 

\section{Physical and Optical Properties}

\begin{table}[t]
\squeezetable
\caption{The neutron capture cross section for the various species in \per\, the number of species in solution
relative to Pb for an 80\% solution concentration and the neutron capture rates relative to hydrogen.}
\label{RelNumber}
\begin{tabular}{c d d  d }
Isotope       & neutron Capture     &    Relative Number   & Relative neutron\\ 
              &  Cross Section (b)  &     Density          & Capture Rates \\ 
\hline
Pb            &                     &   1.      &      \\
\nuc{204}{Pb} & 0.70                &   0.01    & 0    \\
\nuc{206}{Pb} & 0.03                &   0.24    & 0    \\
\nuc{207}{Pb} & 0.70                &   0.22    & 0.04 \\
\nuc{208}{Pb} & 0.02                &   0.52    & 0    \\
Cl            &                     &    2.     &      \\
\nuc{35}{Cl}  & 44.00               &   1.52    & 18.3 \\
\nuc{37}{Cl}  & 0.43                &    0.48   & 0.06 \\
O             &                     &   13.6    & 0     \\
\nuc{16}{O}   & 0.00                &    13.57  & 0    \\
\nuc{17}{O}   & 0.23                &    0.01   & 0    \\
\nuc{18}{O}   & 0.00                &    0.03   & 0    \\
H             & 0.32                &    11.3   & 1    \\
e$^-$         & N.A.                &   234.    & N.A.
\end{tabular}
\normalsize
\end{table}

To build a reasonably large detector viewed by photo-multiplier tubes from the
periphery, the attenuation of the $\check{C}$erenkov light must be minimal. Therefore
the transmission of light through the medium is critical. To understand the transmission,
 measurements were made at several wavelengths through several concentrations of \per \cite{DOE99}.
The maximum achievable transmission is still under study with attentuation lengths exceeding 
2.5 m having been measured in an 80\% solution. A large mass detector may require segmentation
to minimize adverse effects from light loss.

The $\check{C}$erenkov response of the medium depends on the index of refraction, which
 for \per\ is 1.5 for an 80\% solution
by weight. The density of this solution is 2.7 g/cm$^3$, the corresponding \nuc{208}{Pb} number density
is $1.7 \times 10^{21}$/cm$^3$ and the hydrogen number density is $3.6 \times 10^{22}$/cm$^3$.
Some additional pertinent data  for \per\ are given in Table \ref{RelNumber}.

The number of $\check{C}$erenkov photons emitted by an energetic electron in \per\
is about 185/cm\cite{PDB98} in the wavelength region of interest
for phototubes. The stopping power of  80\% (50\%) \per\ is about 0.2 cm/MeV (0.33 cm/MeV). 
Hence a 15 MeV electron  will emit a total of about 550 (920) photons in 80\% (50\%) \per . Due
to the modest light yield, the resolution ($\approx$20\%) will be modest. The resonance widths will also
contribute an uncertainty in relating the electron energy to the incoming neutrino energy by a comparable 
amount (few MeV).

Because solutions of \per\ contain a great deal of water and Cl, the neutron thermalization and capture time
is of the order of 10-100 $\mu$s. This time scale is short compared to the $\nu$
detection rate during a supernova, but long enough to identify whether there are 0, 1,
or 2 neutrons in coincidence with a primary e$^-$ or $\gamma$.

\section{Cross Section Models}

The FHM\cite{FUL99} paper pointed out the importance of the \nuc{208}{Pb}
 - $\nu$ cross section and 
the production of neutrons. They noted that the total number of neutrons produced by neutrino 
interactions with Pb is sensitive to the effective temperature of the supernova $\nu_e$
 energy spectrum especially for reactions which produce multiple neutrons.
 However, they did not consider detection schemes
 that could measure the energy released in the form of electrons and gamma rays (hereafter 
referred to as the electromagnetic energy). Neutral current events release a large number of neutrons
but very little electromagnetic energy. Hence 
NC events can be isolated from CC events by measuring the electromagnetic energy
 in coincidence with the neutrons.    

The papers by FHM and KL provided effective cross sections for neutrino-lead 
interactions averaged over the neutrino energy spectra and summed over all the 
product nucleus states. It is just these distributions, however, which are needed to simulate
 the response of a detector and to optimize its design parameters. This section summarizes the 
nuclear physics model implemented to calculate these distributions. In the following section, these 
distributions and their use in analyzing the neutrino spectra are described.

\begin{table}[t]
\squeezetable
\caption{A description of the levels used in the simulation. The 2 neutron rate from the NC
 1st forbidden transition is not critical to the analysis. The neutron number branching
ratios used here were motivated by results from FHM.}
\label{NucLevels}
\begin{tabular}{ c c c c c  }
Transition        & Product Nucleus        &  Fraction        & Fraction         & Fraction          \\ 
                  & Energy Level           &   decays          & decays           & decays            \\ 
                  & w.r.t. Pb              &   with 0 n        & with 1 n         & with 2 n           \\ 
\hline
CC IAS            &  17.53 MeV              & 0    & 1    & 0     \\
CC GT             &  17.9 MeV               & 0    & 0.9  & 0.1    \\
CC first forb.    &  24.1 MeV               & 0    & 0    & 1    \\
NC GT             &  7.32 MeV               & 0.55 & 0.45 & 0   \\
NC first forb.    &  14 MeV                 & 0    & $\sim$ 1  & $\sim$ 0 
\end{tabular}
\normalsize
\end{table}

Specifically, the energy dependence of the five primary transitions of 
interest as identified by FHM is calculated: two for the NC interaction and three for the CC interaction. The parameters
 describing these transitions are given in Table \ref{NucLevels}. For the isobaric analog state (IAS) CC
and Gamow-Teller (GT) NC and CC transitions, the analytic expressions and matrix element
values from Ref. \cite{FUL99} are used. For the first 
forbidden transitions, the cross section can not be written in closed form.
Software to calculate that contribution to the cross section as used by FHM was obtained\cite{HAX00} and hence the
cross section calculations presented here agree with FHM by design. 

\begin{figure}
\begin{center}
\includegraphics[width=3.375in]{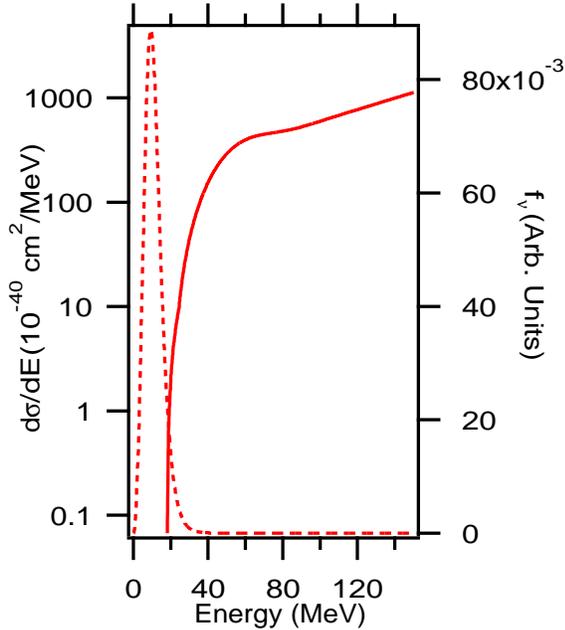}
\end{center}
\caption{The CC cross section as a function of energy plotted on the left axis (solid line). 
A normalized thermal neutrino energy spectrum (T$_{\nu_e}$ = 2.76 MeV, $\eta$ = 3.) is plotted on 
the right axis (dashed line). Note the cross section scale is logarithmic whereas the spectrum scale
is linear.}
\label{CrossSection3}
\end{figure}

\begin{figure}
\begin{center}
\includegraphics[width=3.375in]{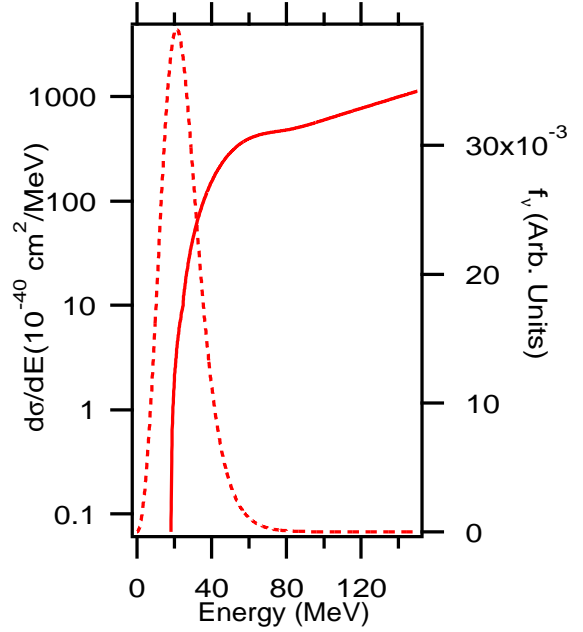}
\end{center}
\caption{The CC cross section as a function of energy plotted on the left axis (solid line). 
A normalized thermal neutrino energy spectrum (T$_{\nu_e}$ = 6.27 MeV, $\eta$ = 3.) is plotted on 
the right axis (dashed line). Note the cross section scale is logarithmic whereas the spectrum scale
is linear.}
\label{CrossSection8}
\end{figure}

Figure 2 shows the cross section overlaid on a normalized thermal flux distribution ($T_{\nu_e}$=2.76
MeV, $\eta$=3). Figure 3 shows the same for a higher neutrino temperature ($T_{\nu_e}$=6.27
MeV, $\eta$=3). The origin of the large increase in the interaction rate with temperature is clear from
a comparison of 
these plots. The onset of the Pb cross section lies between these two temperature extremes. As a
result, the interaction rate in Pb is highly sensitive to $T_{\nu_e}$. This is especially true for 
the first forbidden transition which leads to the emission of 2 neutrons.

\begin{table}[t]
\squeezetable
\caption{Cross sections for $\nu_e$ interactions on 
several isotopes, for various energy spectra 
 and authors. The column labeled by DAR indicates an average of the cross section over
the muon decay-at-rest energy spectrum for electron neutrinos. The column labeled SN is
for a Fermi-Dirac distribution of energies with $\eta$ = 3 and $T_{\nu}$ =
6.266 MeV ($\overline{E}_{\nu}$ $\sim$ 25 MeV). The cross sections are in
units of $10^{-40}$ cm$^2$. For reference various other cross sections are given.}
\label{CrossSection}
\begin{tabular}{c c c c }
Isotope                                  & At 30 MeV         & DAR                  & SN                                       \\ 
\hline
\nuc{208}{Pb}($\nu_e,e^-$)X              & 31. \cite{KOL99a}, 44\tablenotemark[1] & 59.6\cite{KOL00}     &   34.\cite{KOL00}, 47.5\cite{FUL99}    \\
\nuc{37}{Cl}($\nu_e,e^-$)\nuc{37}{Ar}    & 1.0 \cite{KUR90}  &                      &                                           \\
\nuc{35}{Cl}($\nu_e,e^-$)X               &                   &                      &    0.62\cite{FUL99}                       \\
\nuc{35}{Cl}($\bar{\nu}_{e},e^+$)X       &                   &                      &    0.14\cite{FUL99}                       \\
\nuc{16}{O}($\nu_e,e^-$)X                & 0.01 \cite{HAX88} &                      &   0.11\tablenotemark[2] \cite{HAX88}      \\
\nuc{16}{O}($\nu_x,\nu_x'$)X             &                   &                      &   0.03\cite{KOL99a}                       \\
D($\nu_e,e^-$)X                          & 0.41 \cite{KUB94} & 0.53 \cite{YIN92}    &   0.4\cite{KUB94}                         \\
H($\bar{\nu}_{e},e^+$)n                  & 0.6\cite{VOG84}   &                      &   0.5                                     \\
$\nu_e$(e,e')$\nu_e'$                    & 0.003 \cite{BAH89}&                      &   0.0023\tablenotemark[2] \cite{HAX88}      
\end{tabular}
\tablenotetext[1]{This value was calculated by the author using the formalism of Ref. {\protect \cite{FUL99}}.}
\tablenotetext[2]{This value is for {\protect $\eta$ = 0 and $T_{\nu}$ = 8.0 MeV.}}
\normalsize
\end{table}

Tables \ref{CrossSection} and \ref{RelInteraction} show some cross sections and
 relative interaction rates for the various
components of \per . It is clear that the interactions with Pb and H dominate.
 Contributions from Cl, O, and elastic scattering
of electrons are at the few 
percent level. The CC interaction
of  $\bar{\nu}_{e}$ on protons is a major contribution because the hydrogen
density is large in \per\ solutions. For other isotopes found in \per\ but not listed in the table, 
FHM note that the cross section models vary like N-Z or NZ/A. For
Pb, these dependencies are small for the naturally occurring isotopes. Hence one
can anticipate that their response to neutrinos will be similar. This is not
the case for Cl. However, the difference in cross section between \nuc{35}{Cl} and \nuc{37}{Cl}
is still probably smaller than other uncertainties and thus the Cl interaction
rate is small.

\begin{table}[t]
\squeezetable
\caption{The neutrino interaction rates for 30 MeV $\nu_e$ for the dominant reactions.
The values are given relative to \nuc{208}{Pb}.
The cross section value for Pb corresponds to that of FHM.}
\label{RelInteraction}
\begin{tabular}{ c d }
Element        &    Relative Rate          \\ 
\hline
\nuc{208}{Pb}($\nu_e$,e)                  &    100.     \\
\nuc{37}{Cl}($\nu_e$,e)                   &      2.0    \\
\nuc{16}{O}($\nu_e$,e)                    &      0.6    \\
H($\bar{\nu}_{e},e^+$)n                   &     32.     \\
$\nu_e$(e,e')$\nu_e'$                     &      3.1
\end{tabular}
\normalsize
\end{table}

\section{The Electromagnetic Energy}

A detector based on
lead perchlorate would be able to measure the energy deposited by e$^-$
and $\gamma$ in coincidence with the number of neutrons produced. 
The energy deposited in the detector due to e$^-$ and $\gamma$ is here 
referred to as the electromagnetic energy to contrast it with that resulting from 
neutron capture. To identify a particular class of neutrinos,
one sorts the electromagnetic energy spectra by the number of coincident neutrons.
CC events always produce an electron which can have substantial energy
 in coincidence with neutrons. NC events
produce either a high-energy $\gamma$ with no neutrons or neutrons with little or no
$\gamma$ energy.  

First consider the separation of NC from CC events. The product 
nucleus from the neutrino interaction can produce 0, 1, or 2 neutrons. (Particle decays 
other than neutrons (alphas and protons) of the product 
nucleus states have a much lower branching ratio (see KL) and are ignored here.) When the
interaction does 
produce a neutron, the neutrons carry away the available energy and 
 there is little electromagnetic energy produced. Hence for the NC events, the
electromagnetic energy is only appreciable when 0 neutrons are emitted. Thus events which do have appreciable
electromagnetic energy in coincidence with neutrons are indicative of
\nuc{208}{Pb} - $\nu_e$ or H - $\bar{\nu}_e$ CC interactions.

\begin{figure}
\begin{center}
\includegraphics[width=3.375in]{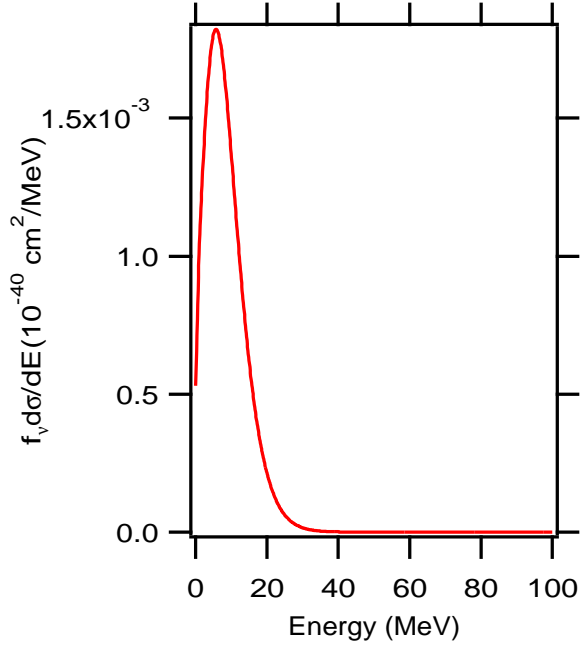}
\end{center}
\caption{The energy observed in the detector due to electrons from the CC interaction that are 
found in coincidence with 2 neutrons ($E_2$). This plot is for T$_{\nu_e}$ = 2.76 MeV, $\eta$ = 3.}
\label{TwoSuper3}
\end{figure}

\begin{figure}
\begin{center}
\includegraphics[width=3.375in]{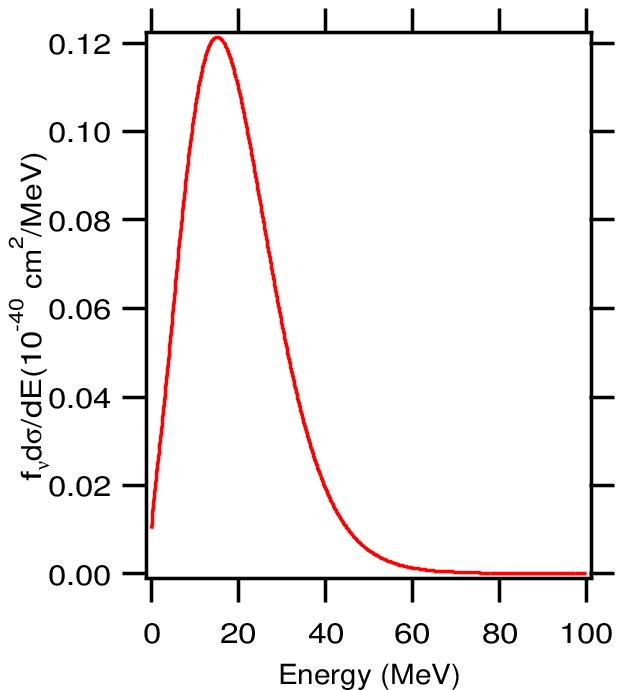}
\end{center}
\caption{The energy observed in the detector due to electrons from the CC interaction that are 
found in coincidence with 2 neutrons ($E_2$). This plot is for T$_{\nu_e}$ = 6.27 MeV, $\eta$ = 3.}
\label{TwoSuper8}
\end{figure}

\begin{figure}
\begin{center}
\includegraphics[width=3.375in]{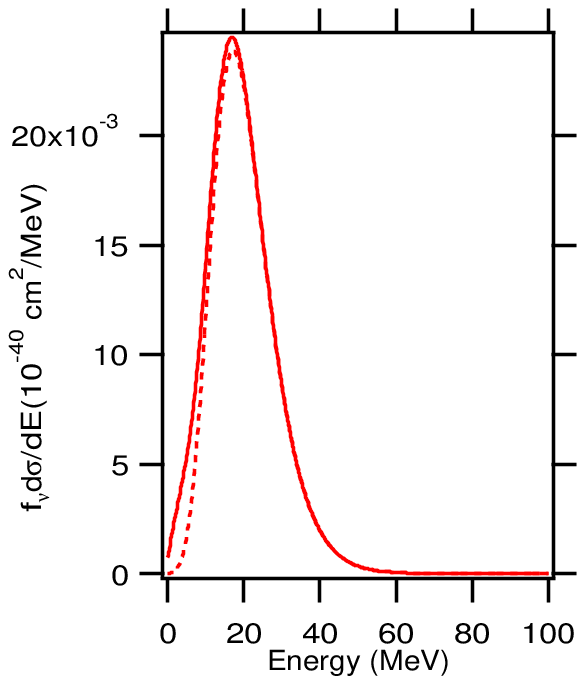}
\end{center}
\caption{The energy observed in the detector due to electrons from the CC interaction that are 
found in coincidence with 1 neutron ($E_1$). This plot is for T$_{\nu_e}$ = 2.76 MeV, 
T$_{\bar{\nu}_e}$ = 4.01 MeV, $\eta$ = 3.
The dashed line represents the contribution due to $\bar{\nu}_e$ on H.}
\label{OneSuper3}
\end{figure}

\begin{figure}
\begin{center}
\includegraphics[width=3.375in]{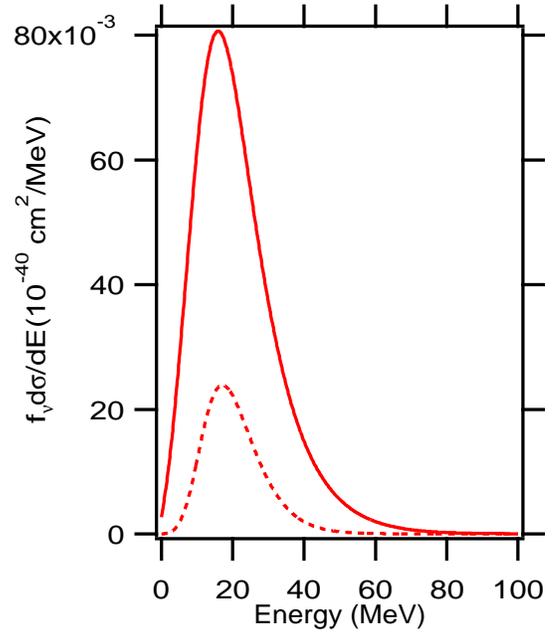}
\end{center}
\caption{The energy observed in the detector due to electrons from the CC interaction that are 
found in coincidence with 1 neutron ($E_1$). This plot is for T$_{\nu_e}$ = 6.27 MeV, 
T$_{\bar{\nu}_e}$ = 4.01 MeV, $\eta$ = 3.
The dashed line represents the contribution due to $\bar{\nu}_e$ on H.}
\label{OneSuper8}
\end{figure}

Next consider the separation of the CC events into $\nu_e$ or $\bar{\nu}_e$ events.
The 2-neutron event rate is
 dominated by $\nu_e$-induced  CC transitions to the first forbidden
 level in Bi. Since the electromagnetic energy for the CC reactions is related to the incoming 
neutrino energy, the 2-neutron event spectrum provides data on the incident $\nu_e$ energy
 spectrum. In contrast, the IAS and GT transitions induced by $\nu_e$ 
produce mostly 1-neutron events. Furthermore, although $\bar{\nu}_e$ do not have an appreciable
 CC cross section on Pb, they have a large cross section on H. 
The 1-neutron spectrum is therefore generated by reactions induced by $\nu_e$ 
and $\bar{\nu}_e$. But those are the only reactions that contribute significantly to the 1-neutron
 electromagnetic energy spectrum and therefore the 1-neutron spectrum provides data on the $\bar{\nu}_e$
 energy distribution. The 2-neutron and 1-neutron spectra thus  relate directly to the spectral features
of the incident neutrino flux. Figs. \ref{TwoSuper3} through \ref{OneSuper8} show spectra of the
electromagnetic energy for events in coincidence with 1 or 2 neutrons for two different
neutrino energy distributions. These spectra have been convolved with an estimate for the detector
resolution.

For NC events, only GT excitations can produce an
 event with no neutrons as the resonance straddles the neutron separation energy.
 A $\gamma$ ray of approximately 7.3 MeV is then emitted. This
 would produce a line in the 0-neutron electromagnetic energy spectrum. NC transitions can also produce
 1-neutron events and a modest number of 2-neutron events. As there will be little
electromagnetic energy in coincidence, the 1-neutron events will be interpreted as an 8.6-MeV
line in the 0-neutron spectrum. The strengths of these two lines in the 0-neutron spectrum provide a
measure of the NC rate. Due to the resolution, these two lines will most likely be
blended. These rates will be large even for low-temperature profiles.

One background to the CC 1-neutron spectrum comes from 
2-neutron NC events. One of the 2 neutrons would be taken as a neutron but the second
would be mistaken for electromagnetic energy. For $T_{\nu_{\mu},\nu_{\tau}}$=6.27, $T_{\nu_e}$
=2.76, $\eta$=3, the NC rate of 2-neutrons events will be comparable to the CC rate\cite{FUL99,KOL00} because,
although the NC cross section is much smaller, there are 4 times as many $\nu_{\mu,\tau}$ as $\nu_e$
and their temperature is much higher. Thus at low $T_{\nu_e}$, this background must be separated. 
Note that because there are several $\gamma$ sharing the energy and
 photons Compton scatter more than once before losing all their energy,
the $\check{C}$erenkov light produced by neutrons capturing on Cl will be isotropic in
comparison to electrons from inverse beta decay. This could lead to a technique for separating
this background. Otherwise an electromagnetic energy threshold of 10 MeV would eliminate it.

\begin{figure}
\begin{center}
\includegraphics[width=3.375in]{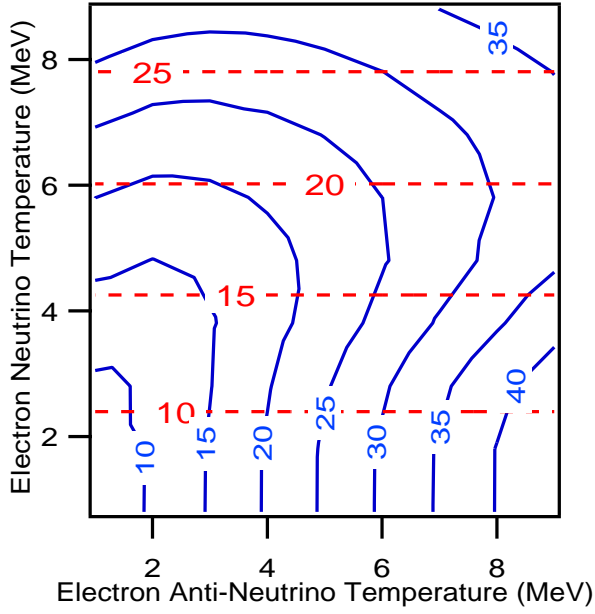}
\end{center}
\caption{A contour plot of \EOne\ (solid lines) and \ETwo\ (dashed lines) as a function of the neutrino
 temperatures. In both cases, the average energy was calculated for events depositing at least 5 MeV.}
\label{TempContour}
\end{figure}

\begin{figure}
\begin{center}
\includegraphics[width=3.375in]{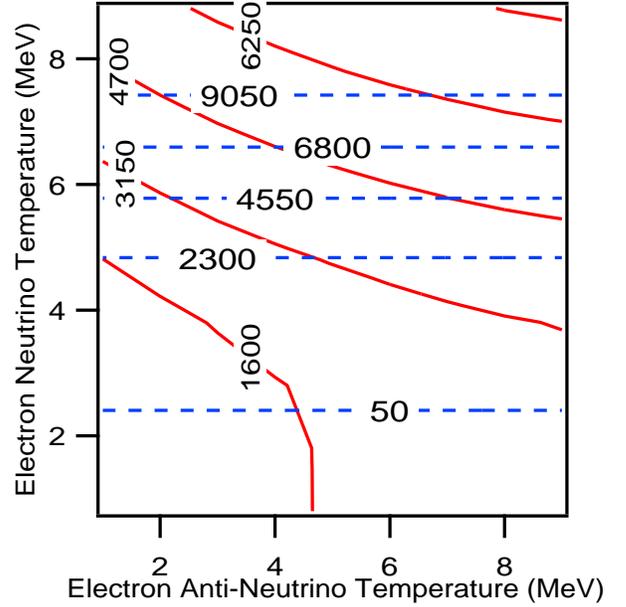}
\end{center}
\caption{This is a contour plot of the count rates as a function of the neutrino temperature 
for a 24 kt \per\ detector. The solid lines
are for the number of 1-neutron CC events and the dashed lines are for the number of 2-neutron CC events.
 A threshold of 5 MeV was used. 
}
\label{CountRateContour}
\end{figure}

With the spectra in hand, one deduces the $\nu$ temperature by relating it to 
the average electromagnetic energy from the 1-neutron spectrum (\EOne )  and the
 2-neutron spectrum (\ETwo ) (Fig. \ref{TempContour}).  One also counts the total number of 1-neutron events ($N_1$) and
 2-neutron events ($N_2$) in these spectra (Fig. \ref{CountRateContour}).
 The ratio (R=$N_2$/$N_1$) depends
 strongly on the temperatures of the two types of neutrinos but is insensitive to uncertainties
in the supernova distance scale. Using these
  parameters (\EOne , \ETwo , $N_2$, $N_1$, R), the pair of neutrino temperatures that fit the data are determined.

If the electron neutrino temperature is clearly greater than that of the electron
 anti-neutrino temperature, it would be considered strong evidence for $\nu_e-\nu_{\mu}$ or $\nu_e-\nu_{\tau}$
 oscillations. This condition is satisfied by the top left
 hand section of these two-temperature plots. It should be noted that the contour plots were done for thermal spectra.
If $\nu$ oscillate, the spectral shape could be complicated depending on the oscillation conditions
and this analysis would have to be extended.

\begin{figure}
\begin{center}
\includegraphics[width=3.375in]{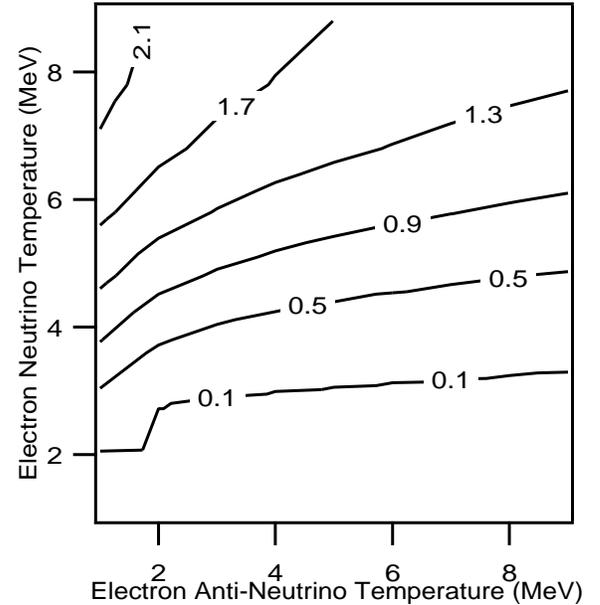}
\end{center}
\caption{This a contour plot of R as a function of the neutrino temperatures. R is 
the ratio of two-neutron events to one-neutron events. A threshold of 5 MeV was used. 
}
\label{RatioContour}
\end{figure}

To use this set of figures to determine the temperatures with some precision, one must estimate the
 uncertainty in the measured parameters. These uncertainties are mostly determined by the number
 of events observed. Regions in Figs. \ref{TempContour} through \ref{RatioContour} are defined by the values
of these parameters and their uncertainties. The boundaries of these regions indicated 
on the axes provide the corresponding temperature ranges that describe the data. For a commonly considered
 supernova at 10 kpc with $3\times10^{53}$ ergs of 
released energy and a 10 kt-Pb detector (24 kt of lead perchlorate), one expects to  measure
 \ETwo\ = 11.0 $\pm$ 0.5 MeV, \EOne\ = 19.6 $\pm$ 0.2 MeV, $N_1$ = 1530, $N_2$ = 82, and R = 0.053 $\pm$ 0.006  
for non oscillating spectra ($T_{\nu_e}$ = 2.76, $\eta$ = 3, $\bar{E}_{\nu_e}$ = 11; 
$T_{\bar{\nu}_e}$ = 4.01, $\eta$ = 3, $\bar{E}_{\bar{\nu}_e}$ = 16). The uncertainties for the 
average energies were estimated by calculating the variance of the distribution and dividing by
the square root of the number of observed counts.
 The resulting statistical uncertainty in the temperatures is small and the precision
improves if $T_{\nu_e}$ is larger because the event rate is higher.

\section{Summary}

The Sudbury Neutrino Observatory (SNO)\cite{SNO} and Super-Kamiokande\cite{SK} experiments are operating.  In the event
of a supernova, Super-Kamiokande
will observe a large number of $\bar{\nu}_e$-H events ($\sim 10000$ for $T_{\bar{\nu}_e}$ = 4.01, $\eta$ = 3) and SNO
will see a number of $\nu_e$-D events ($\sim 90$ for $T_{\nu_e}$ = 2.76, $\eta$ = 3). The average
energy measured for these two reactions could be compared in a two-temperature plot in a similar analysis as
described above. Due to the smaller number of events, the
uncertainty comparing these energy distributions would be determined by the SNO measurement. For these
$\nu$ temperatures, a 24 kt
\per\  experiment would count a number of 2-neutron $\nu_e$ events comparable to the number of
 SNO D($\nu_e$,e$^-$)pp events. Thus
the average energy analysis of a \per\ experiment would have a sensitivity similar to the SNO-SK comparison. 
But the total number of counts in a \per\ experiment is much more sensitive to the neutrino temperatures. SNO (\per\ ) would
observe about 3 (60) times as many D($\nu_e$,e$^-$)pp (2-neutron) events at $T_{\nu_e}$ = 6 than at 2.76 MeV.
This extra handle on the data adds a powerful redundancy to the analysis.

It is obvious from Table \ref{CrossSectionComparison} that there are still large uncertainties in the nuclear 
physics of these interactions. Furthermore, reactions on the other Pb isotopes have not been considered although they would
tend to increase the signal.
 Therefore the response of such a detector should be measured
 in a prototype. The $\nu_e$ spectrum from stopped $\mu^+$ decay is comparable to
 that from supernovae. So the 2-neutron and 1-neutron spectra could be measured at a beam stop
facility \cite{ORLaND}
 without the influence of $\bar{\nu}_e$. Since the $\bar{\nu}_{e}$-proton
 reaction is well understood, it does not need additional study and the
 measurement would focus on what is currently poorly known: the nuclear physics parameters
 of the lead reactions. Thus the theoretical uncertainties in the cross section could be
diminished by a beam stop experiment. 

The details of the analysis presented in this paper will alter as the nuclear
 physics of this system becomes better known. But the general conclusions generic to lead
detectors will not. It is very powerful to measure not just the total
number of neutrons produced but the 2-neutron and 1-neutron rates separately.
Furthermore the observation of electrons and gamma rays in coincidence with the neutrons can permit the
separation of NC and CC interactions. If one observes a supernova with a Pb detector,
 these combinations of data can be analyzed to
estimate the temperatures of the neutrino spectra.

\begin{table}[t]
\squeezetable
\caption{Some cross section comparisons between FHM and KL. 
The cross sections are given in units of $10^{-40}$ cm$^2$.
 In most cases the two calculations have fair agreement. The selection
here was chosen to emphasize the differences.}
\label{CrossSectionComparison}
\begin{tabular}{ c c c }
Transition and                                    & FHM     &  KL          \\ 
spectral input                                    &         &              \\ 
\hline
Pb($\nu_e$,e)X   $T_{\nu_e}$=4.0 MeV, $\eta$=3.   &   6.9   & 6.7    \\
Pb($\nu_e$,e)X   $T_{\nu_e}$=8.0 MeV, $\eta$=0.   &   58    & 43    \\
Pb($\nu$,$\nu$')X $T_{\nu_e}$=4.0 MeV, $\eta$=3.  &   0.66  & 0.23    \\
Pb($\nu$,$\nu$')X $T_{\nu_e}$=8.0 MeV, $\eta$=0.  &   4.5   & 1.4   \\
Pb($\nu$,e)X  $\pi^+$ decay at rest               &  91\tablenotemark[1]   & 59.6 
\end{tabular}
\tablenotetext[1]{This value was calculated by the author using the formalism of FHM.}
\normalsize
\end{table}

\section{Acknowledgements}

I would like to thank Peter Doe, Hamish Robertson, Thomas Steiger, John Beacom,
 Edwin Kolbe, Gail McLaughlin and George Fuller for useful conversations.
I am especially grateful to Wick Haxton for providing code to estimate the first forbidden
transition cross section. 
This research was support by the Department of Energy,
grant number DE-FG03-97ER41020.


\begin{thebibliography}{99}

\bibitem{LAND} C. K. Hargrove, {\it et al.}, Astropart. Phys. {\bf 5}, 183 (1996).
\bibitem{OMNIS} D. B. Cline {\it et al.}, Phys. Rev. {\bf D50}, 720 (1994); 
P. F. Smith, Astropart. Phys. {\bf 8}, 27 (1997).
\bibitem{HAR98} C. K. Hargrove, private communication.
\bibitem{FUL99} George M. Fuller, Wick C. Haxton, and Gail C. McLaughlin, Phys. Rev. {\bf D59}, 085005 (1999).
\bibitem{KOL99a} E. Kolbe, private communication at "Low-Energy Neutrino Physics" program at the
Institute for Nuclear Theory (INT-99-2), Summer 1999.
\bibitem{KOL00} E. Kolbe and K. Langanke, Preprint Nucl-th/0003060.
\bibitem{CRC} Handbook of Chemistry and Physics, 65$^{th}$ edition, ed. R. C. West, (CRC Press, Boca Raton, FL, 1984).
\bibitem{MIL93} D. S. Miller, J. R. Wilson, and R. W. Mayle, Astrophys. J. {\bf 415}, 278 (1993).
\bibitem{JAN89} H.-T. Janka and W. Hillebrant, Astron. Astrophys. J. {\bf 224}, 49 (1989).
\bibitem{DOE99} P. J. Doe, S. R. Elliott, C. Paul, and R. G. H. Robertson, Nucl. Phys. {\bf B} (Proc. Suppl.) {\bf 87}, 512 (2000).
\bibitem{PDB98} C. Caso, {\it et al.}, Europ. Phys. J. {\bf 3}, 1 (1998).
\bibitem{KUR90} T. Kuramoto, M. Fukugita, Y. Kohyama, and K. Kubodera, Nucl. Phys. {\bf A512}, 711, (1990).
\bibitem{HAX88} W. C. Haxton, Nucl. Instrum. Meth. {\bf A264}, 37, (1988).
\bibitem{KUB94} K. Kubodera and S. Nozawa, Inter. J. Mod. Phys. 3, 101 (1994).
\bibitem{YIN92} S. Ying, W.C. Haxton, and E. M. Henley, Phys. Rev. {\bf C45}, 1982 (1992).
\bibitem{VOG84} P. Vogel, Phys. Rev. {\bf D29}, 1918 (1984), and P. Vogel and J. F. Beacom, PR {\bf D60}, 053003 (1999).
\bibitem{BAH89} John N. Bahcall, {\it Neutrino Astrophysics}, (Cambridge U. P., Cambridge, 1989).
\bibitem{HAX00} W. C. Haxton, private communication.
\bibitem{SNO} J. Boger, {\it et al.}, Nucl. Instrum. and Meth. {\bf A449}, 172 (2000).
\bibitem{SK} Y. Fukuda, {\it et al.}, Phys. Rev. Lett. {\bf 82}, 2430, (1999).
\bibitem{ORLaND}  F. T. Avignone {\it et al.} (the ORLaND Collaboration), in Perspectives in Nuclear Physics, eds J.
H. Hamilton, H. K. Carter, and R. B. Piercy, World Scientific, 1999, p 25-34.

\end{thebibliography}
\end{document}